\documentclass[prd,eqsecnum,nofootinbib,twocolumn,letterpaper]{revtex4}

\usepackage{fbb} 
\usepackage[libertine,bigdelims,vvarbb]{newtxmath}
\usepackage[cal=boondoxo]{mathalfa}
\usepackage[T1]{fontenc}
\useosf

\usepackage{bm,graphicx,mathrsfs}
\usepackage{amssymb,amsmath}
\usepackage[ugly]{units}
\usepackage[colorlinks=true,citecolor=blue,bookmarks=true]{hyperref}

\newcommand{\bx}{\bm{x}}
\newcommand{\bg}{\bm{g}}
\newcommand{\bz}{\bm{z}}

\newcommand{\bA}{\bm{A}}

\newcommand{\bF}{\bm{F}}
\newcommand{\bS}{\bm{S}}
\newcommand{\bL}{\bm{L}}
\newcommand{\rmd}{\mathrm{d}}

\newcommand{\Ept}{ E_\mathrm{pt} }
\newcommand{\rc}{ r_\mathrm{circ} }

\begin{document}
\title{Extended-body effects and rocket-free orbital maneuvering}

\author{Abraham I. Harte}

\author{Michael T. Gaffney}

\affiliation{Centre for Astrophysics and Relativity, School of Mathematical Sciences
	\\
	Dublin City University, Glasnevin, Dublin 9, Ireland}

\begin{abstract}
	The trajectory of a spherical object which falls freely in a gravitational field is fixed by its initial position and velocity. However, an object which can control its shape can also control its motion: Except where forbidden by symmetries and their associated conservation laws, a shape-changing (but rocket-free) spacecraft can have complete control over its trajectory. We discuss a general formalism which allows rocket-free maneuvers to be understood without constructing detailed interior models. A spacecraft's interior is abstracted to the specification of a quadrupole moment, and in some cases, it is only a single eigenvalue of that moment which is relevant. For orbits around a spherically-symmetric mass, we show that appropriately varying the relevant eigenvalue allows the energy and eccentricity of an orbit to be increased or decreased and its apsides to be rotated arbitrarily. Strategies are identified which optimize these maneuvers. In other contexts, we show that extended-body effects can be used to stabilize orbits which would otherwise be unstable.

\end{abstract}

\maketitle

\vskip 2pc

\section{Introduction}

How can a spacecraft change its orbit? The obvious answer is to use a rocket. But there is another possibility: It can generically ``grab onto'' or ``push off of'' gradients in a gravitational field, effectively using them as invisible, omnipresent handholds. More precisely, bodies can control the gravitational forces and torques which act upon them by manipulating their internal mass distributions. Although the resulting extended-body effects typically perturb gravitational forces only slightly from their point-particle counterparts, the integrated consequences of these perturbations can be considerable; large, fuel-free maneuvers are possible for those who wait.

Examples of such maneuvers have been discussed before, typically for spacecraft consisting of pairs of tethered masses in orbit around spherically-symmetric bodies \cite{MITpaper, Landis, Beletsky, Gratus, Longo, Burov, Murakami}. In those cases, it has been shown that cyclically reeling a tether in and out can result in secular changes to an orbit's energy, eccentricity, and apsidal orientation.

Explicit models such as tethered masses have the advantage of being conceptually concrete. However, they obscure the underlying physics. It is unclear which results are common to all spacecraft capable of rocket-free maneuvers and which are artifacts of the particular model which was considered. Here, we explain how extended-body effects can be understood more broadly, without adopting any particular model of a spacecraft's interior. Our method allows all of the internal dynamics to be abstracted to the specification of a spacecraft's quadrupole moment tensor. With this in hand, it is straightforward to compute which variations in that tensor can be used to execute a desired maneuver. The remaining engineering problem would then be to design an explicit mechanism which allows the quadrupole moment to vary according to the resulting specification. The abstraction of our approach provides a better understanding of what is possible and what is not. It also illustrates that much of the behavior of a shape-changing spacecraft can be understood using simple, intuitive concepts such as the conservation of (a suitably-defined notion of) energy.

Our discussion is initially general but then specializes. Sect. \ref{Sect:Review} considers the motion of essentially-arbitrary extended bodies in arbitrary gravitational fields, deriving the constraints imposed by symmetries (if any), forces and torques in the quadrupole approximation, and certain energies which are relevant in this context. These results are applied in Sect. \ref{Sect:2body} to understand how to alter orbits around spherically-symmetric masses. We specialize to spacecraft which experience no gravitational torque, showing that appropriate cyclic variations in one of the eigenvalues of the quadrupole moment can be used to independently vary an orbit's eccentricity and apsidal orientation. Sect. \ref{Sect:Unstable} considers a second application of the general formalism, where extended-body effects are used not to change an orbit, but to maintain one; unstable orbits are stabilized. It is shown that a shape-changing spacecraft can stabilize itself in a nearly-circular, ordinarily-unstable orbit around a highly-oblate central mass.

\section{Motion of extended bodies}
\label{Sect:Review}

The general theory of extended-body effects in Newtonian gravity is not new, and reviews may be found in, e.g., \cite{PoissonWill, 300years}. To briefly summarize, net forces and torques can be approximated using multipole expansions, and these expansions demonstrate that a body's center-of-mass motion couples only weakly to its internal dynamics. Such results have been applied mainly to astrophysical problems in which the multipole moments of celestial bodies are determined by their rotation and by the tides raised by external gravitational fields. That specialization considerably constrains the types of extended-body effects which may arise, and qualitatively-different behavior is possible for spacecraft which actively manipulate their internal configurations.

This section reviews some aspects of the theory of extended-body motion, allowing for arbitrarily-complicated gravitational fields. Some new results are also discussed regarding conservation laws and their relations to symmetries, and also notions of mechanical energy which are relevant for shape-changing spacecraft.

An extended body with mass $m$ can be characterized in part by its center-of-mass position $\bz(t)$ and its spin angular momentum $\bS(t)$; these are the only characteristics which are constrained by universal laws of physics (as opposed to details of an object's internal composition). To review, these satisfy $m \ddot{\bz} = \bF$ and $\dot{\bS} = \bm{\tau}$, where the net gravitational force $\bF(t)$ and torque $\bm{\tau}(t)$ depend only on the external gravitational field $\bg(\bx,t)$ and on the mass density $\rho(\bx,t)$:
\begin{equation}
	\bF = \int \rho \bg \,\rmd^3 \bx, \qquad \bm{\tau} = \int \rho ( \bx - \bz)  \times \bg \,\rmd^3 \bx.
	\label{FInt}
\end{equation}
Newton's law of gravitation implies that no matter the source(s) of the gravitational field, 
\begin{equation}
	\nabla^2 \bg = 0
	\label{Harmonic}
\end{equation}
throughout the body of interest; the gravitational field is harmonic.

If a body is spherically-symmetric in the sense that $\rho(\bx, t) = \tilde{\rho} (|\bx - \bz(t)|,t)$ for some radial mass density $\tilde{\rho}$, the mean-value property of harmonic functions \cite{MathPhys} can be applied to \eqref{FInt} to show that the torque vanishes and that the force depends only on the gravitational field at the center of mass:
\begin{equation}
    \bF(t) = m \bg (\bz(t),t), \qquad \bm{\tau}(t) = \bm{0}.
    \label{Fsph}
\end{equation}
Although these results are well-known for point particles, they are in fact valid for spherically-symmetric bodies of arbitrary size. It is nevertheless crucial for the discussion below that forces and torques which act on less symmetric mass distributions can be quite different. These differences may be exploited to allow an object to control its motion.

\subsection{What is (im)possible?}
\label{Sect:ConsLaws}

Before describing which forces and torques \textit{can} be produced by appropriately-designed spacecraft, we first discuss those which \textit{cannot}. It may be expected, for example, that if the gravitational field is completely uniform, no internal mechanism could be used to make the force and torque deviate from \eqref{Fsph}; there are no gradients to ``grab on to.'' If the gradient of a gravitational field instead vanishes only in one direction, it is only the component of the force in that direction which cannot be controlled. Generalizing further, we now show that if the gravitational gradient is symmetric in some way, a certain combination of force and torque components cannot differ from their point-particle counterparts. In many (but not all) cases, such a symmetry further implies that a certain combination of linear and angular momentum components must be conserved. 

We begin by considering symmetries of at least the geometry. More specifically, we consider the Killing vector fields \cite{Wald} of three-dimensional Euclidean space, which are the generators of translations and rotations. Letting $\mathcal{T}_i$ denote the components of an arbitrary constant vector and $\mathcal{R}_{ij} = - \mathcal{R}_{ji}$ the components of an arbitrary, constant, antisymmetric rank-2 tensor, all Euclidean Killing vectors are given by\footnote{Here and below, the Einstein summation convention has been applied so that repeated indices are to be summed over all possible values. For example, $\mathcal{R}_{ij} x^j = \sum_{j=1}^3 \mathcal{R}_{ij} x^j$.}
\begin{equation}
    \xi_i( \bx ) = \mathcal{T}_i + \mathcal{R}_{ij} x^j.
    \label{Killing}
\end{equation}
These make up all solutions to Killing's equation $\partial_i \xi_j + \partial_j \xi_i = 0$. They span a six-dimensional space---three dimensions may be ascribed to translations and three to rotations.

Interesting dynamical statements are associated with vector fields which are symmetries not only of the geometry, but also of the gravitational field. To understand this, it is useful to first introduce the ``generalized momentum'' 
\begin{equation}
    P_{\bm{\xi}} (t) \equiv m \dot{\bz}(t) \cdot \bm{\xi}(\bz(t)) + \tfrac{1}{2}  \bS(t) \cdot \left[ \bm{\nabla} \times \bm{\xi}(\bz(t))  \right]
    \label{PDef}
\end{equation}
which is associated with each Killing vector $\bm{\xi}$. This is a linear combination of linear and angular momentum components. All of the linear and angular momenta may in fact be understood as different aspects of the generalized momentum, a point of view developed in more general contexts in \cite{HarteSyms, SFrev}. Regardless, \eqref{FInt} implies that the rate of change of $P_{\bm{\xi}}$, the ``generalized force,'' is
\begin{align}
	\dot{P}_{\bm{\xi}} & =\bF \cdot \bm{\xi} + \tfrac{1}{2} \bm{\tau} \cdot \left( \bm{\nabla} \times \bm{\xi} \right)  
	\nonumber
	\\
	&= \int \rho (\bm{\xi} \cdot \bg) \, \rmd^3 \bx.
	\label{FNgen}
\end{align}
The first line here expands the generalized force as a combination of ordinary force and torque components. The second line determines that combination in terms of a body's mass distribution and the external gravitational field. If a potential $\Phi$ is introduced such that $\bm{g} = - \bm{\nabla} \Phi$, the integrand is seen to be proportional to the Lie derivative \cite{Wald} $\mathcal{L}_{\bm{\xi}} \Phi = ( \bm{\xi} \cdot \bm{\nabla}) \Phi$. It follows that $\dot{P}_{\bm{\xi}} $ may be viewed as a mass-weighted average of the degree by which $\bm{\xi}$ fails to generate a symmetry of $\Phi$. 

We now ask when it is impossible for $\dot{P}_{\bm{\xi}}$ to be controlled by a  mechanism internal to the object of interest: When must the generalized force be equal to its point-particle value $m (\bm{\xi} \cdot \bg)$ for all physically-realizable mass distributions, i.e., for all $\rho \geq 0$ with fixed $m$ and $\bz$? Use of \eqref{FNgen} shows that this occurs when 
\begin{equation}
    \partial_i \partial_j \mathcal{L}_{\bm{\xi}} \Phi = 0.
    \label{symCriterion}
\end{equation}
Any Killing vector which satisfies this criterion may be said to generate a symmetry of the gravitational field. For each such symmetry, a particular combination of force and torque components cannot be controlled. That combination cannot deviate from its point-particle counterpart, meaning that
\begin{align}
	(\bF - m \bg) \cdot \bm{\xi} + \tfrac{1}{2} \bm{\tau} \cdot ( \bm{\nabla} \times \bm{\xi})  = 0.
	\label{FNcons}
\end{align}
This does not necessarily imply a simple conservation law. However, 
\begin{align}
	 P_{\bm{\xi}} = \mathrm{constant}
	 \label{ConsLaw}
\end{align}
if \eqref{symCriterion} is strengthened to $\mathcal{L}_{\bm{\xi}} \Phi = 0$. These results provide the main constraints to what can be accomplished by changing a body's mass distribution. 

Their simplest application is to an extended body in a constant nonzero gravitational field. In that case, all Killing vector fields satisfy \eqref{symCriterion} and it follows from \eqref{FNcons} that the force and torque reduce to \eqref{Fsph} for all mass distributions, not only spherical ones. As expected, extended-body effects cannot be used to control the force and torque in the absence of a gravitational gradient. It may also be noted that in this case, $\bm{g}$ itself is a Killing field, and although $P_{\bm{g}}$ is not conserved (because $\mathcal{L}_{\bm{g}} \Phi \neq 0$), its first derivative is.

It is more interesting to consider gravitational fields which are not constant, but are instead spherically-symmetric. Such fields are preserved by all rotations about the center of the gravitating mass, and choosing this center to be at the origin of the coordinate system, the symmetries can be explicitly represented by all Killing vectors without a translation component: $\mathcal{T}_i = 0$ in \eqref{Killing}. Substituting the resulting vectors into \eqref{FNcons} while varying over all possible $\mathcal{R}_{ij}$ shows that the force determines the torque via 
\begin{equation}
	\bm{\tau} = \bF \times \bz.
	\label{tauSph}
\end{equation}	
Recalling \eqref{PDef} and \eqref{ConsLaw}, it also follows that the angular momentum about the origin must be conserved. More precisely, using $\bL \equiv m \bz \times \dot{\bz}$ to denote the orbital angular momentum about the origin,
\begin{equation}
    \bL + \bS = \mathrm{constant}.
    \label{SphericalLtot}
\end{equation}
It is therefore impossible to use extended-body effects to modify $\bL$ without also modifying $\bS$. This observation has of course been made before, although the discussion here places places it in a wider context and allows similar results to be obtained very easily in different settings. Motion in spherically-symmetric gravitational fields is considered in detail in Sect. \ref{Sect:2body} below.

It can also be interesting to consider motion around a mass which is axisymmetric but not spherically symmetric. There would then be only one symmetry of the gravitational field, namely that associated with rotations around the axis of symmetry. This implies that only the component of $\bL+\bS$ along the symmetry axis would necessarily be conserved. The two remaining components of $\bL$ could therefore be controlled independently of the corresponding components of $\bS$. This freedom  might be used to engineer a spacecraft capable of controlling its orbital inclination without spinning itself up. A different aspect of motion in axisymmetric gravitational fields is briefly considered in Sect. \ref{Sect:Unstable} below.

\subsection{The quadrupole approximation}

Now that we have outlined what is not possible, we discuss some of what is. In principle, \eqref{FInt} may be used to compute corrections to the point-particle force and torque for general extended bodies. Directly evaluating the relevant integrals would appear to require detailed knowledge of a body's internal mass distribution. This is rarely necessary, however. Useful approximations arise when the gravitational field varies only slightly inside the object of interest, which occurs whenever its size is small compared to the distances to other masses. In these cases, corrections to the point-particle forces and torques may be described using a only a handful of parameters which characterize some aspects of the spacecraft's internal state. Crucially, these parameters do not characterize that state completely; most of the information contained in $\rho$ is irrelevant to the dynamics and may be ignored.

Assuming that $\bg$ does vary slowly inside the body of interest, it may be expanded for points near the center of mass. Substituting the first three terms of such an expansion into the force integral results in
\begin{equation}
	F_i = m g_i + \tfrac{1}{2} Q^{jk} \partial_j \partial_k g_i + \ldots,
	\label{Fexpand}
\end{equation}
where the field and its derivatives are to be evaluated here at $\bz$ and 
\begin{equation}
	Q^{ij} \equiv \int  \rho \left[ ( x^i - z^i ) (x^j - z^j) - \tfrac{1}{3} \delta^{ij} |\bx - \bz|^2 \right]\rmd^3 \bx
	\label{QDef}
\end{equation}
denotes the components of the spacecraft's quadrupole moment. This moment is symmetric and trace-free and may depend on time. It follows from \eqref{Harmonic} that any term proportional to $\delta^{jk}$ could be added to $Q^{jk}$ in \eqref{Fexpand} without affecting $F_i$, and this freedom has been used in \eqref{QDef} to ensure that the quadrupole moment is always trace-free. It may be related to the inertia tensor $I_{ij}$ via $Q^{ij} = - (\delta^{ik} \delta^{jl} - \frac{1}{3} \delta^{ij} \delta^{kl} ) I_{kl}$. Unlike the inertia tensor, the quadrupole moment tensor vanishes when $\rho$ is spherically symmetric. 

Crucially, there are no general laws of physics which control the time dependence of $Q^{ij}$. A suitably-engineered spacecraft can vary its quadrupole moment essentially at will, as long as it remains symmetric and trace-free. The only restriction is that the magnitudes of its various components must be bounded above by approximately $m \ell^2$, where $\ell$ denotes the spacecraft's maximum spatial extent. Rather than adopting explicit models of deformable spacecraft, it is convenient to instead abstract their engineering to the specification of $Q^{ij}$. This has five independent components, which may be viewed as five control parameters. If the gravitational field is generated by a single massive body at distance $r \equiv |\bz|$, the quadrupole terms in the force are suppressed by a factor of order $(\ell/r)^2$ relative to the leading-order monopole force $|m \bm{g}|$. The omitted terms in \eqref{Fexpand} are suppressed by an additional factor of $\ell/r$. 

A body's quadrupole moment affects not only the forces which act upon it, but also the torques. Using \eqref{FInt} together with $\bm{\nabla} \times \bg = \bm{0}$ shows that in an arbitrary gravitational field,
\begin{equation}
	\tau_i= \epsilon_{ijk} Q^{jl} \partial_l g^k + \ldots,
	\label{tauexpand}
\end{equation}
where $\epsilon_{ijk}$ denotes the permutation symbol [so $(\bm{a} \times \bm{b})_i = \epsilon_{ijk} a^j b^k$ for any vectors $\bm{a}$ and $\bm{b}$]. In the context described in the previous paragraph, the magnitude of this torque is of order $(\ell/r) (| \bm{F} | \ell)$. The omitted terms in \eqref{tauexpand} are suppressed relative to this by an additional factor of $\ell/r$.

All discussion below is confined to the quadrupole approximation. Ellipses and other notations which denote approximation are nevertheless omitted for brevity.

\subsection{Energy}
\label{Sect:energy}

We now discuss the energy of an extended body. Except in special circumstances, the mechanical energy of a shape-changing spacecraft---meaning the sum of its kinetic and gravitational potential energies---is not conserved. This is because non-mechanical reservoirs of energy, such as batteries, may be needed in order to perform the work required to rearrange masses in a nonzero tidal field. Taking into account all relevant types of energy would require a detailed internal model, and even if such a model were available, the resulting total would not necessarily be useful for understanding orbital dynamics. Nevertheless, certain model-independent notions of mechanical energy are useful for understanding motion. We consider two.

The first of these is the ``point-particle energy'' which would be conserved in a static field and in the absence of extended-body effects. This energy is defined via
\begin{equation}
    \Ept \equiv \tfrac{1}{2} m |\dot{\bz}|^2 + m \Phi,
    \label{EDef2}
\end{equation}
a definition which may be applied regardless of whether or not a body actually is pointlike. Allowing for time variation and quadrupole effects,  \eqref{Fexpand} implies that
\begin{equation}
    \dot{E}_\mathrm{pt} = m \partial_t \Phi + \tfrac{1}{2} \dot{z}^i Q^{jk} \partial_i \partial_j  g_k.
\end{equation}
Although this energy is not generically conserved, it can be useful. It describes the mechanical energy a spacecraft could have---as defined in a conventional sense---if it were to rapidly rearrange itself into a configuration for which $Q^{ij} = 0$. In this sense, it might be viewed as a kind of ``osculating energy.''

The second notion of energy we consider differs by incorporating all of the gravitational potential energy
\begin{equation}
	U = \int \rho \Phi \, \rmd^3 \bx,
	\label{Uint}
\end{equation}
rather than only its point-particle approximation $m \Phi$; let
\begin{equation}
	E \equiv \tfrac{1}{2} m |\dot{\bz}|^2 + U.
	\label{EDef}
\end{equation}
This still excludes any internal contributions to the kinetic energy, as well as, e.g., thermal energy and non-gravitational reservoirs of potential energy.

Regardless, expanding $\Phi$ near $\bz$ in \eqref{Uint} shows that the two energies here are related via
\begin{equation}
    E = \Ept - \tfrac{1}{2} Q^{ij} \partial_i g_j
    \label{Ediff}
\end{equation}
in the quadrupole approximation. Moreover,
\begin{equation}
	\dot{E} = \left[  m \partial_t \Phi - \tfrac{1}{2} Q^{ij} \partial_t (\partial_i g_j) \right] - \tfrac{1}{2} \dot{ Q }^{ij} \partial_i g_j.
	\label{Edot}
\end{equation}
The first group of terms here vanishes when the gravitational field is static, and in those cases, $E$ \textit{is conserved whenever}
\begin{equation}
    \dot{ Q }^{ij} \partial_i g_j = 0.
    \label{QdotZero}
\end{equation}
If quadrupole moments are changed only stepwise at discrete times in a static field---which is an idealization of many useful control strategies---the conservation of $E$ can be used in between each configuration change. Furthermore, changes in $E$ from one step to the next are given by
\begin{equation}
	\delta E = \delta U = - \tfrac{1}{2} \delta Q^{ij} \partial_i g_j  . 
	\label{dEGen}
\end{equation}
This contrasts with the point-particle energy, for which $\delta \Ept = 0$ after a rapid configuration change. Most orbital maneuvers considered below involve spacecraft whose quadrupole moments satisfy \eqref{QdotZero} except at certain discrete times where an internal mechanism is used to switch between different configurations.

\section{Orbits around a spherical mass}
\label{Sect:2body}

We now apply the general formalism developed in Sect. \ref{Sect:Review} to analyze the motion of a shape-changing spacecraft with mass $m$ and quadrupole moment $Q^{ij}$ which is in orbit around a spherically-symmetric body with mass $M \gg m$. Again defining $r = |\bz|$, the gravitational potential at the spacecraft's center of mass is
\begin{equation}
	\Phi = - \frac{ GM }{ r }.
	\label{gPoint}
\end{equation}
While motion in this potential has been analyzed before \cite{MITpaper, Landis, Beletsky, Gratus, Longo, Burov, Murakami} for certain tethered spacecraft, we now discuss it from a perspective in which spacecraft are characterized only by their quadrupole moments.

Evaluating the gradient of the potential while using \eqref{Fexpand} and defining $\hat{\bz} \equiv \bz/r$, the force in the quadrupole approximation is seen to be
\begin{equation}
	F_i = m g_i + \frac{ 3 G M }{ r^4 } \left[ Q_{ij} \hat{z}^j - \tfrac{5}{2} ( Q_{jk} \hat{z}^j \hat{z}^k) \hat{z}_i \right].
	\label{Fpoint}
\end{equation}
Similarly, \eqref{tauexpand} implies that
\begin{equation}
	\tau_i = - \frac{ 3 G M }{ r^3 } \epsilon_{ijk} \hat{z}^j ( Q^{kl} \hat{z}_l ) . 
	\label{FNpoint}
\end{equation}
The force and torque thus depend on the quadrupole moment only in the combination $Q^{ij} \hat{z}_j$.

\subsection{Torque-free spacecraft}
\label{Sect:torquefree}

As discussed in Sect. \ref{Sect:ConsLaws}, the spherical symmetry of the gravitational field implies that the total angular momentum about the origin must be conserved. A rocket-free spacecraft attempting to change $\bL$ can therefore do so only at the cost of spinning itself up. However, the degree of spin-up required to \textit{significantly} change $\bL$ would typically be so large as to result in structural failure.

Consider, for example, a spacecraft consisting of two equal parts connected by a retractable massless tether with maximum length $\ell$. If such a spacecraft begins in a circular orbit with radius $r$ and controls its tether---using, perhaps, the strategy described in \cite{Gratus}---in order to eventually bring itself to a new circular orbit with the slightly-different radius $r + \Delta r$, the centripetal accelerations of the end-masses due to the consequent rotation would be of order
\begin{equation}
%     \frac{ 2 G M }{ r^2 } 
     2 |\bg|\left( r / \ell \right)^3 \left( \Delta r / r \right)^2 
\end{equation}
at full extension and much more as the tether is reeled in. Even a $1\%$ increase in radius from low-Earth orbit would thus induce accelerations of order $(\unit[5 \times 10^{5}]{m/s^2}) (\unit[10]{km}/ \ell)^3$. It is therefore impractical to use extended-body effects to meaningfully change $\bL$ in most contexts\footnote{This assumes orbits around large celestial bodies and interactions which are purely gravitational. It may be possible in practice to remove excess angular momentum by interacting with a planet's magnetic field or with the solar wind \cite{Murakami}. For orbits around smaller objects such as comets or asteroids, even this may not be necessary \cite{Landis}.}. This precludes certain orbital maneuvers. However, it also illustrates that extended-body effects can be used to control a spacecraft's spin without significantly altering its orbit.

If the intent is to control an orbit and not a spin, and only small changes in $\bL$ or $\bS$ can reasonably be produced, the analysis may be simplified by restricting to spacecraft in which these vectors do not change at all. Then $\bm{\tau} = \bm{0}$ and \eqref{tauSph} implies that $\bF \propto \hat{\bz}$; the direction of the force cannot be controlled in this context. Torque-free extended bodies may nevertheless control the magnitude of the gravitational force which acts upon them. To understand what the lack of direction control excludes, first recall that for an elliptical Keplerian orbit with semi-major axis $a$ and eccentricity $e$,
\begin{equation}
    L \equiv |\bL| = m \sqrt{GM a (1-e^2)}.
    \label{L}
\end{equation}
A process which conserves $L$ may thus increase $a$ but only by increasing $e$ as well. A torque-free spacecraft cannot control these parameters independently. Note in particular such spacecraft cannot use extended-body effects to move from one circular orbit to another.

Regardless, the design of a torque-free spacecraft can be abstracted to the specification of a single control parameter. To see this, first note that it follows from \eqref{FNpoint} that the torque vanishes if and only if the radial vector is an eigenvector of the quadrupole moment. We thus require that
\begin{equation}
	Q^{ij} \hat{z}_j = q \hat{z}^i
	\label{eigQ}
\end{equation}
for some eigenvalue $q$. The sign of $q$ is unconstrained in general, as its time dependence. It is the relevant control for torque-free spacecraft in spherically-symmetric gravitational fields. In terms of it, \eqref{Fpoint} implies that the force is
\begin{equation}
	\bF = \left[ 1 + \tfrac{ 9 }{ 2 } (q/m r^2) \right] m \bg.
	\label{Fsimp}
\end{equation}

As a simple example in which the eigenvector condition \eqref{eigQ} is satisfied, consider again a spacecraft which consists of two equal parts connected by a massless strut with length $\ell$. If the strut is perpendicular to the orbital plane, \eqref{QDef} implies that $q = - m \ell^2/12$. If the strut is oriented radially, $q = m \ell^2/6$. The sign and magnitude of $q$ may thus be controlled by varying the length and orientation of the strut---the latter perhaps with the help of a reaction wheel or similar mechanism. Of course, these are only examples. Our analysis is not restricted to these two types of spacecraft.

Regardless of internal details, we now suppose that $q$ can be engineered to be piecewise constant. This is an idealization, although it is one which allows the orbital dynamics to be understood very easily. In particular, \eqref{QdotZero} is satisfied when $\dot{q}=0$, implying that $E$ is conserved whenever $q$ is constant. The conservation of $L$ additionally allows the angular contribution to the kinetic energy to be rewritten so \eqref{EDef} reduces to
\begin{equation}
    E = \tfrac{1}{2} m\dot{r}^2 + U_\mathrm{eff}(r,q),
    \label{Eeff}
\end{equation}
where 
\begin{equation}
    U_\mathrm{eff}(r,q) \equiv \frac{ L^2 }{ 2 m r^2 } + \left[ 1 +\tfrac{3}{2} (q/mr^2) \right] m \Phi .
    \label{Ueff}
\end{equation}
The first term in $U_\mathrm{eff}$ is the centrifugal potential energy and  the remaining terms are equal to $U$. The conservation of $E$ for fixed $q$ implies that the radial motion is equivalent to the motion of a point particle in only one dimension with effective potential energy $U_\mathrm{eff}$.

If $q$ rapidly switches from one fixed value to another at $r = r_{\mathrm{sw}}$, both $U_\mathrm{eff}$ and $E$ change. It follows from \eqref{dEGen} that
\begin{equation}
	\delta E = \delta U = \delta U_{\mathrm{eff}} = \tfrac{3}{2}\big( \delta q/ m r_{\mathrm{sw}}^2 \big) m\Phi  .
	\label{deltaE1}
\end{equation}
The change in energy thus depends on the radius at which a spacecraft rearranges its mass distribution. Larger jumps occur at smaller radii where the tidal gradient is larger. A cycle in which $q$ increases once when $r$ is small and returns to its original value when $r$ is large may be seen to lead to a net decrease in $E$. Reversing the order here would instead result in a net increase in $E$. Regardless, if $q$ is cycled many times in such a manner, the energy of an orbit can be considerably altered without the use of a rocket. This is not unlike the process by which a child uses cyclic motion to pump up a swing.

\subsection{Characterizing orbits}

A bound, torque-free spacecraft in which $q$ is fixed and small for all time has an orbit which may be described as very close to a slowly-precessing ellipse. Its radii of pericenter $r_-$ and apocenter $r_+$ may be found, for example, by solving $U_\mathrm{eff}(r_\pm,q) = E$. This equation is time independent so solutions are fixed and orbits do not raise or fall over time. 

The dynamics can be quite different when $q$ varies. Appropriately-timed cyclic changes in this parameter can be used to slowly transition from one nearly-elliptical orbit to another, quite-different orbit. It is convenient to describe such processes using the Laplace-Runge-Lenz vector \cite{Goldstein}
\begin{equation}
    \bA \equiv \frac{ \dot{\bz} \times \bL }{ GM m }    - \frac{ \bz }{ r },
    \label{Adef}
\end{equation}
sometimes also referred to as the eccentricity vector. For the $q=0$ Kepler problem, this is known to be constant.  In that case, it is directed from the origin to the pericenter of the orbit and its magnitude is equal to the eccentricity. If $q$ is nonzero, $\bA$ is not constant. Still, its magnitude may be used to \textit{define} an ``instantaneous eccentricity'' via
\begin{equation}
	e \equiv | \bA |.
	\label{eDef}
\end{equation}
Similarly, the direction of $\bA$ may be interpreted as pointing from the origin towards an orbit's ``instantaneous pericenter.'' To make this precise, first define the true anomaly $\nu$ to be the angle of the spacecraft away from its instantaneous pericenter so
\begin{equation}
	e \cos \nu = \bA \cdot \hat{\bz}.
\end{equation}
Substituting \eqref{Adef} into the right-hand side of this equation results in
\begin{equation}
    r = \frac{ (L/m)^2 }{ GM } (1+e \cos \nu)^{-1},
    \label{rEllipse}
\end{equation}
which describes the shape of the orbit. If $e$ were constant and $\nu$ the angle away from a fixed direction in space, it would be the standard representation for an ellipse with one focus at the origin, eccentricity $e$, and semi-major axis 
\begin{equation}
	a = \frac{ (L/m)^2 }{ GM (1-e^2) }.
	\label{aDef}
\end{equation}
In the systems considered here, $e$ varies slowly and $\nu$ is an angle away from the slowly-varying direction tangent to $\bA$. Eq. \eqref{rEllipse} may thus be interpreted as describing a continuously-varying sequence of ellipses. The ``instantaneous semi-major axis lengths'' of these ellipses may be defined using \eqref{aDef}, which preserves the Keplerian relation \eqref{L} between $L$, $a$, and $e$.

If $\nu$ varies much more rapidly than $e$, the pericenter and apocenter are approximately given by the Keplerian expressions $r_\pm = a (1 \pm e)$. Using \eqref{EDef2} shows that with this approximation, $a$ can be written as
\begin{equation}
	a = - \frac{ GM m }{ 2 \Ept}
	\label{aDef2}
\end{equation}
at least at the apsides. Moreover, the conservation of $L$ implies that $a (1-e^2) = r_+ r_- /a $ is preserved. The geometric average of an orbit's maximum and minimum radii thus varies in proportion with $(-\Ept)^{-1/2}$.

\subsection{Controlling energy}
\label{Sect:ecc}

\begin{figure}
    \centering
    \includegraphics[width=1\linewidth]{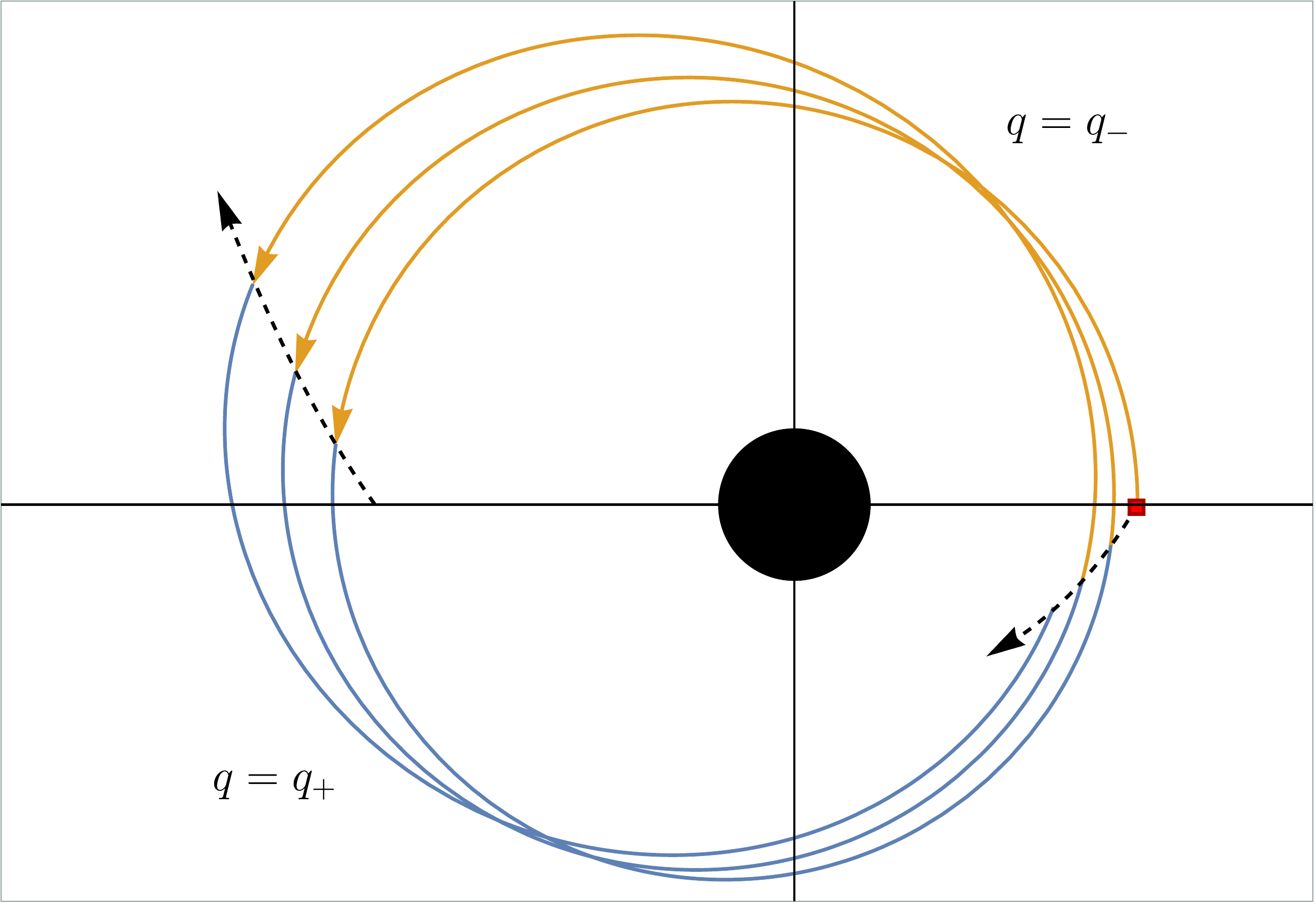}
    \caption{Numerical solution for three orbits with the eccentricity-raising strategy in which $q$ is given by \eqref{qEcc}. In the orange portions of the orbits, $q = q_- = -m a_0^2/100$. In the blue portions, $q = q_+ = 0$. The initial eccentricity is $e_0 = 1/10$. The spacecraft is orbiting counter-clockwise in this diagram. Its orbit is precessing in the clockwise (retrograde) direction. The red marker denotes the beginning of the integration. The dotted arrows denote the lines of apocenter of pericenter predicted by the orbit-averaged eccentricity \eqref{deInt} and precession angle \eqref{psiEcc}.}
    \label{Fig:Orbit}
\end{figure}

We now assume that $E < 0$ and consider cyclic changes in $q$ which increase $E$. As remarked at the end of Sect. \ref{Sect:torquefree}, this energy may be increased by supposing that $q$ switches between two discrete values when the spacecraft is at two different radii. If $q_-$ and $q_+ > q_-$ are constants, suppose in particular that
\begin{enumerate}
    \item After passing pericenter, $q = q_-$.
    \item After passing apocenter, $q = q_+$.
\end{enumerate}
Using $\Theta(\cdot)$ to denote the Heaviside step function, this is equivalent to letting
\begin{equation}
	q =  q_- \Theta\left( \dot{r} \right)+  q_+ \Theta \left( -\dot{r} \right) .
	\label{qEcc}
\end{equation}
It follows from \eqref{deltaE1} that with this strategy, $E$ increases at pericenter and decreases at apocenter, remaining constant in between these points. The net change\footnote{Our convention is that $\delta$ denotes a change which occurs in much less time than an orbital period. A change over one complete orbit is instead denoted by $\Delta$.} $\Delta E$ over one cycle---from just after one pericenter to just after the next---is positive. To first order in $q_\pm$, 
\begin{equation}
	\frac{ \Delta E }{ |E | } =  6 e \left( \frac{  q_+ - q_- }{  m a_0^2 } \right) \frac{ 3 + e^2  }{ (1-e_0^2)^3 },
	\label{dE}
\end{equation} 
where $a_0$ and $e_0$ are the initial values of $a$ and $e$. Use of \eqref{Ediff}, \eqref{aDef}, and \eqref{aDef2} shows that with the same approximation, changes in $E$ are related to changes in $\Ept$, $a$, and $e$ via
\begin{equation}
	\frac{ \Delta E }{ |E | } = \frac{ \Delta \Ept }{ |\Ept | } = \frac{\Delta a }{ a } = \frac{ 2 e \Delta e }{ 1-e^2 } .
	\label{de}
\end{equation}
As $(q_+ - q_-)/ m a_0^2$ is of order $(\ell/a_0)^2 \ll 1$, where $\ell$ again denotes the spacecraft's characteristic size, changes in $a$ and $e$ are small over one cycle. Larger changes may be produced by repeating the process many times. Also, for fixed $a_0$, the effectiveness of this method is enhanced for more eccentric orbits. This is to be expected from the larger tidal gradients encountered along orbits with smaller pericenters. Three cycles are illustrated in Fig. \ref{Fig:Orbit}.

\begin{figure}
    \centering
    \includegraphics[width=1\linewidth]{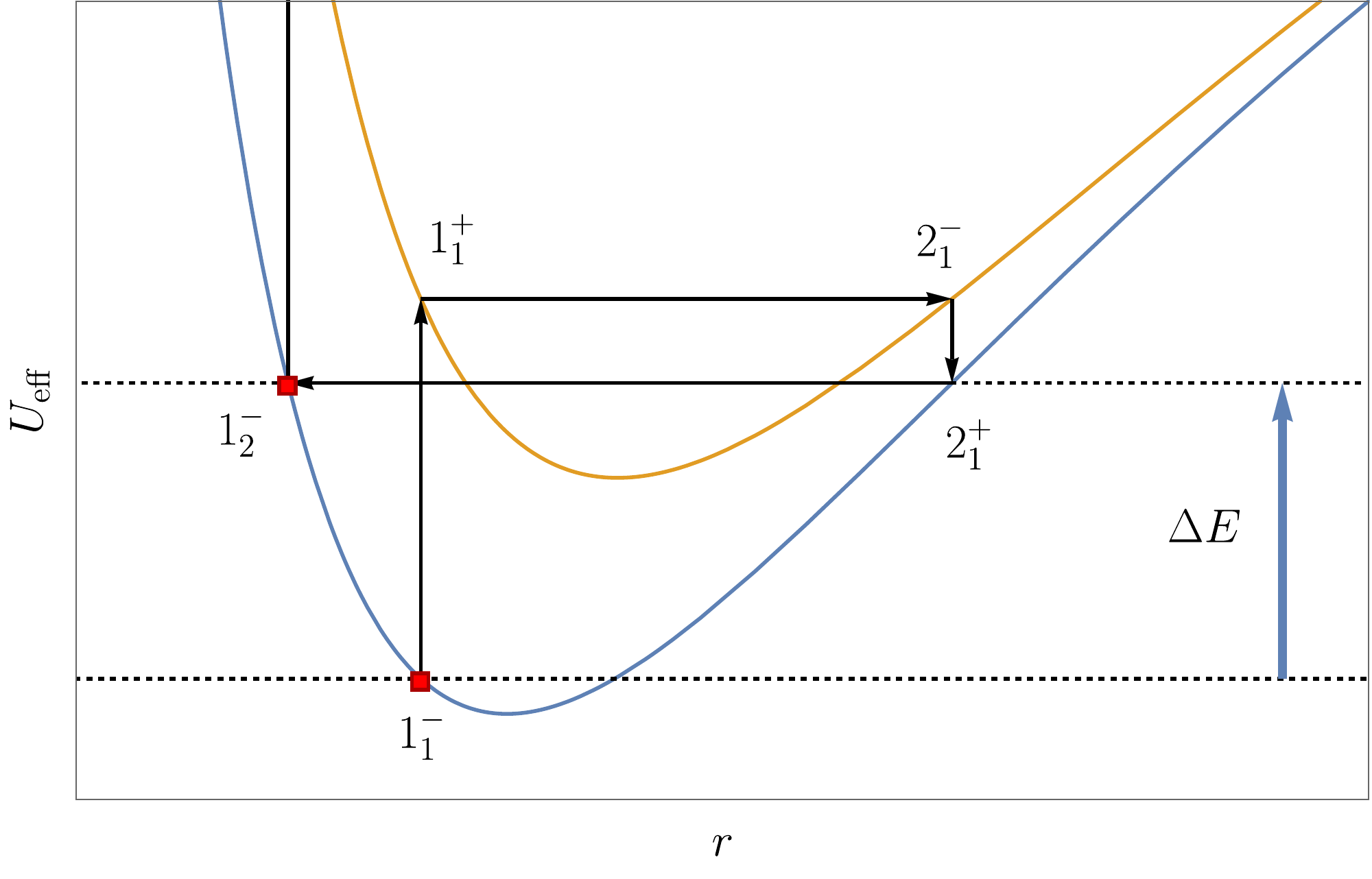}
    \caption{Energy gain using cyclic changes in $q$. The curves are plots of $U_\mathrm{eff}$ with $q = q_+$ (lower) and with $q=q_-$ (upper). In the first cycle, a spacecraft begins in the lower left portion of the diagram just before pericenter, with $q = q_+$ and an energy denoted by the lower dotted line. Upon reaching pericenter, $q$ is decreased to $q_-$. This transition is labeled by the points $1_1^-$ and $1_1^+$. The body's radius then increases until reaching apocenter at the point $2_1^-$. It switches to $q=q_+$ between the points $2_1^-$ and $2_1^+$. Point $1_2^-$ is analogous to $1_1^-$ but represents the moment before switching to $q_-$ in a second cycle which is otherwise omitted. The red squares denote the beginning and end of the first cycle, and $\Delta E$ the energy gain from that cycle. Over many cycles, a spacecraft's energy slowly ``climbs up'' the two effective potential energy curves.}
    \label{Fig:Ueff}
\end{figure}

The consequences of repeatedly cycling $q$ in the given manner may be understood graphically by plotting the effective potential energy \eqref{Ueff}. Unlike in more conventional problems, there are two such curves here: one for $q=q_+$ and one for $q=q_-$. A spacecraft switches between these two states at $r = r_\pm$ and changes in energy which occur during those transitions correspond to differences between the two effective potential energies at those radii. One cycle is illustrated from this perspective in Fig. \ref{Fig:Ueff}. Executing multiple cycles in such a diagram would result in a spacecraft's state oscillating back and forth between the two curves while slowing drifting upwards as $E$ increases.

\begin{figure}
    \centering
    \includegraphics[width=1\linewidth]{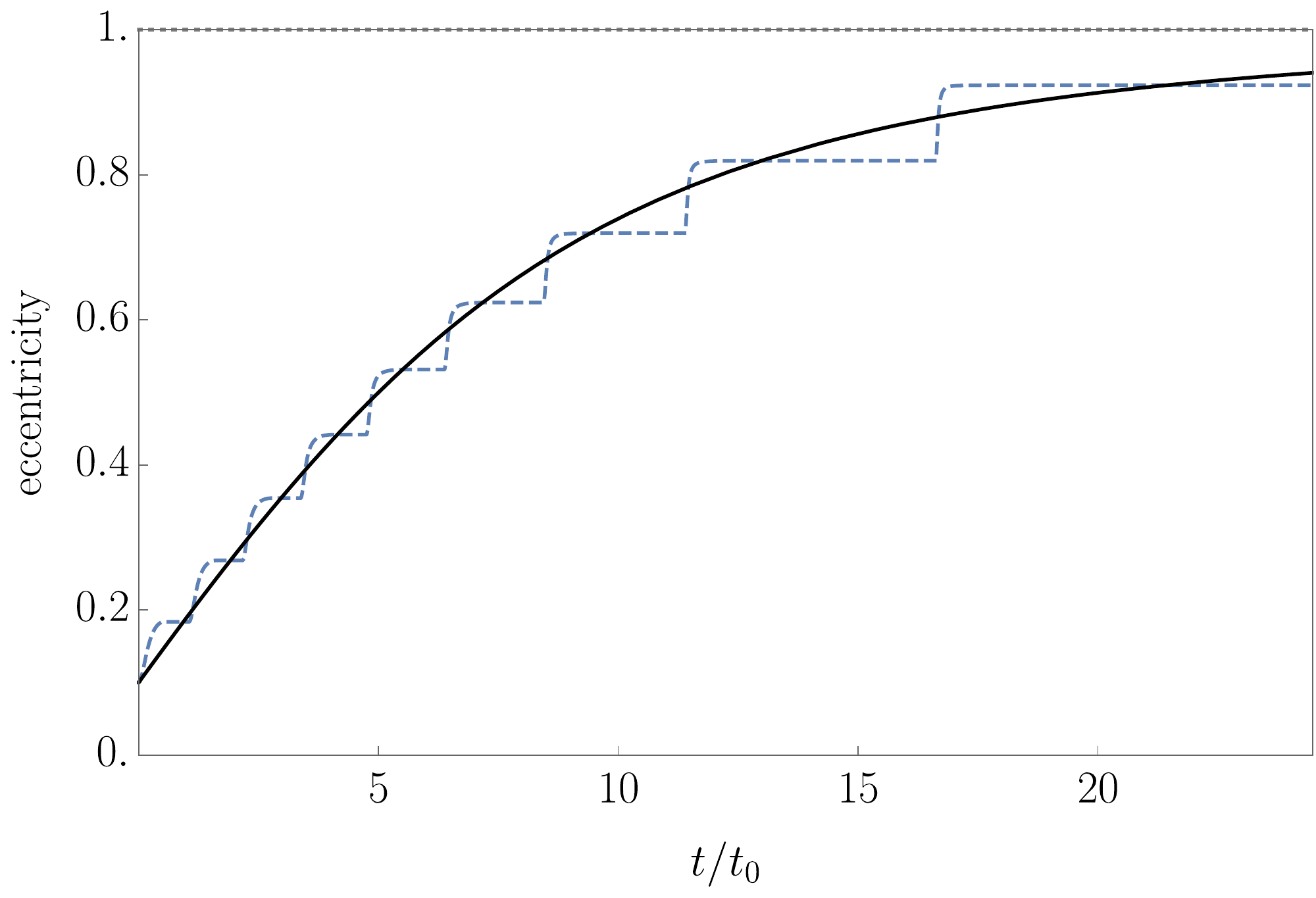}
    \caption{Eccentricity growth using cyclic changes in $q$. The solid curve represents $\langle e \rangle$, which is obtained from \eqref{deInt} together with $\rmd t/\rmd N = 2\pi \langle a \rangle^{3/2} /\sqrt{GM}$ to convert from a function depending on $N$ to one depending on time. The dashed curve represents $e$ and is obtained by direct numerical integration of the equations of motion. $t_0$ denotes the initial orbital period. The values of $e_0$ and $q_\pm$ are the same as in Fig. \ref{Fig:Orbit}.}
    \label{Fig:ecc}
\end{figure}

To be more quantitative about what happens when $q$ is cycled multiple times, the net eccentricity change may be found by repeatedly applying \eqref{dE} and \eqref{de}. Doing so $N$ times determines the orbital parameters just after the $(N+1)$st pericenter. It is however simpler to instead allow $N$ to take non-integer values and to solve the differential equation 
\begin{equation}
	\frac{ d \langle e \rangle }{ dN } =  3 \left( \frac{  q_+ - q_- }{  m a_0^2 } \right) \frac{ 3 + \langle e\rangle ^2  }{ (1-e_0^2)^2 },
	\label{deAv}
\end{equation}
where $\langle e \rangle$ denotes an orbit-averaged eccentricity. This equation is constructed such that $\langle e \rangle$ agrees with $e$ when $N$ is an integer. Defining
\begin{align}
    N_\mathrm{char}^{-1} \equiv \frac{ 3 \sqrt{3}}{ (1-e_0^2)^2 } \left( \frac{  q_+ - q_- }{ m a_0^2 } \right),
    \label{NcInt}
\end{align}
its solution is 
\begin{equation}
   \langle e \rangle = \sqrt{3} \tan \left[ N / N_\mathrm{char} +  \tan^{-1} \left(  e_0 / \sqrt{3} \right)  \right] .
    \label{deInt}
\end{equation}
Similar expressions for orbit averages of $a$ and $\Ept$ may be found by substituting into \eqref{aDef} and \eqref{aDef2}. The orbit-averaged eccentricity is plotted in Fig. \ref{Fig:ecc} together with the true eccentricity obtained by numerically solving the full equations of motion.

These results may be used to compute the number of cycles required to change between any two eccentricities or any two values of $a$ or $\Ept$. At least formally, they may also be used to find the number of cycles required to escape the central body; setting $\langle e \rangle = 1$ gives 
\begin{equation}
    N_\mathrm{esc} = \left[ \frac{\pi}{6} - \tan^{-1} \left( e_0 / \sqrt{3} \right) \right] N_\mathrm{char}.
    \label{Nesc}
\end{equation}
Before an escape could occur, however, the decreasing pericenter would likely in a crash into the central mass. For two equal masses connected by a radially-oriented tether with maximum length $\ell$ and minimum length zero which is initially in a circular, low-Earth orbit, $N_\mathrm{char} = (5 \times 10^5) (\unit[10]{km}/\ell)^2$. Large changes in the orbital parameters thus require either a very large number of cycles or a very long tether.

A spacecraft changing its quadrupole moment in the manner discussed thus far experiences secular increases in its eccentricity and semi-major axis. These parameters may instead be decreased by replacing $\dot{r}$ with $- \dot{r}$ in \eqref{qSwitch}. Such a strategy might be used to, e.g., circularize an initially-eccentric orbit. Angular momentum conservation implies that if an orbit is fully circularized, its final radius would be $\rc = a_0 (1-e_0^2)$. Furthermore, noting that \eqref{deInt} remains valid in this context with the replacement $N_\mathrm{char} \rightarrow -N_\mathrm{char}$, the number of cycles required to eliminate all eccentricity is $N_\mathrm{circ} = N_\mathrm{char} \tan^{-1} (e_0/\sqrt{3})$.

Whether the intention is to increase $e$ or to decrease it, it is natural to ask if the strategies just discussed are as effective as possible, at least for torque-free spacecraft. They are. Energy arguments alone cannot be used to easily verify this. Instead, the eccentricity definition \eqref{eDef} may be directly differentiated while using \eqref{Fsimp} and \eqref{Adef} to yield
\begin{equation}
	\Delta e = - \frac{ 9 }{ 2 } \int_0^{2\pi} \frac{ q }{ ma_0^2} \left( \frac{ 1 + e \cos \nu}{ 1-e_0^2} \right)^2  \sin \nu  d\nu
	\label{deGen}
\end{equation}
to first order in $q$. The sign of the integrand here is equal to the sign of $q \sin \nu$, so if $\Delta e$ is to be as large as possible, $q$ should be as small as possible when $\sin \nu > 0$ and as large as possible when $\sin \nu < 0$. If $q$ is constrained to lie in the interval $[q_-, q_+]$, this results in it being given by \eqref{qEcc}. The strategy presented above is therefore optimal at least for quadrupole moments which are constrained in this way.
 
Eq. \eqref{deGen} may be used to derive changes in the orbital parameters even when $q$ is not piecewise constant [in which case results such as \eqref{deltaE1} are not useful]. This may be understood quite generally by assuming that $q$ depends only on $\nu$ and that it is $2\pi$-periodic. It may then be shown that only a finite number of terms in the Fourier series for $q$ contribute to $\Delta e$. For the best sinusoidal cycle $q = \frac{1}{2} (q_+ - q_-) \sin \nu$, $\Delta e$ is smaller by the factor $(3\pi/16)[1+ 1/(3+e^2)] <1$ than it would have been for an optimal cycle with the same amplitude. For the next most efficient sinusoidal cycle $q = \frac{1}{2} (q_+ - q_-) \sin 2\nu$, the eccentricity change is less than the optimal value by the factor $(3\pi/4) e/(3+e^2) < 1$.

\subsection{Controlling orbital orientation}
\label{Sect:precession}

Besides controlling $a$ and $e$, extended-body effects can also be used to control the orientation of an eccentric orbit in the plane orthogonal to $\bL$. We again assume a torque-free spacecraft whose internal state is parametrized by the quadrupole eigenvalue $q$. If this eigenvalue vanishes, the orbital orientation is fixed. If it is nonzero and constant, Bertrand's theorem \cite{Goldstein} implies that the orbit precesses; its orientation slowly rotates.

To quantify this and also to understand it for a time-dependent $q$, recall that the Laplace-Runge-Lenz vector $\bA$ defined by \eqref{Adef} points from the origin towards the instantaneous pericenter. It changes direction as described by the angular velocity $\bm{\omega} \equiv \bA \times \dot{\bA} / e^2$, so a torque-free spacecraft subject to the force \eqref{Fsimp} precesses at the instantaneous rate
\begin{equation}
    \bm{\omega} = \frac{ q }{ m r^2 }\left( \frac{ 9  \bL }{ 2 em r^2 } \right)  \cos \nu.
    \label{omega}
\end{equation}
Letting $\psi$ denote the total precession angle so $\bm{\omega} = \dot{\psi} \hat{\bm{L}}$, this may be used together with \eqref{rEllipse} to show that the precession angle per orbit is
\begin{equation}
    \Delta \psi = \frac{ 9 }{ 2 e} \int_0^{2\pi} \left( \frac{ q }{ m a_0^2 } \right)  \left( \frac{ 1+e \cos \nu  }{1-e_0^2 } \right)^2  \cos \nu  \rmd \nu .
    \label{dpsiGen}
\end{equation}
Note the similarity between this expression and its analog \eqref{deGen} for $\Delta e$.

If $q$ varies according to the optimal eccentricity-raising cycle \eqref{qEcc}, direct integration shows that
\begin{equation}
    \Delta \psi = \frac{ 9 \pi }{2 (1-e_0^2)^2 } \left( \frac{q_+ + q_- }{ma_0^2} \right).
    \label{psiEcc}
\end{equation}
This is constant, so after $N$ cycles, $\psi = N \Delta \psi$. Also note that $\Delta \psi$ is determined by $q_+ + q_-$ while \eqref{dE} and \eqref{de} imply that $\Delta e$ is determined by $q_+ - q_-$. Appropriate choices for $q_+$ and $q_-$ may therefore be used to control the eccentricity and the precession rate \textit{independently}. Nevertheless, although switching $q$ at apocenter and pericenter maximizes eccentricity changes, it does not maximize precession rates.

To find a strategy which does optimize these rates, note that the sign of the integrand in \eqref{dpsiGen} is given by the sign of $q \cos \nu$. If it is again assumed that $q$ can vary only in the interval $[q_-,q_+]$, the maximum amount of prograde precession is seen to be obtained by setting $q=q_+$ when $\cos \nu > 0$ and $q=q_-$ otherwise. Doing so results in
\begin{align}
    \Delta \psi = \frac{ 9 \pi }{ 2(1-e_0^2)^2 } \Bigg[2 \left( \frac{ q_+ - q_- }{ ma_0^2 }  \right) \left( \frac{1 + 2 e_0^2/3 }{\pi e_0} \right)
    \nonumber\\
    ~ + \left( \frac{ q_+ + q_- }{ ma_0^2 } \right) \Bigg].
    \label{psiOpt}
\end{align}
If a large amount of retrograde precession is desired instead, $q$ may be set equal to $q_-$ when $\cos \nu > 0$ and $q_+$ otherwise. This results in a precession given by \eqref{psiOpt} but with the roles of $q_+$ and $q_-$ reversed. Use of \eqref{deGen} also shows that for both the prograde and retrograde strategies, $\Delta e = 0$. One of their interesting features is that at fixed $a_0$, their effectiveness is enhanced at both small and large eccentricities. While large-eccentricity enhancements also arise in, e.g., \eqref{de}, we find low-eccentricity enhancement only for the precession rate. It is perhaps unsurprising that an orbit which is only slightly eccentric should be relatively simple to reorient.

\section{Stabilizing unstable orbits}
\label{Sect:Unstable}

As another application of the formalism in Sect. \ref{Sect:Review}, we now show that a shape-changing spacecraft can stabilize itself in an otherwise-unstable orbit. This is not of course applicable to motion in the spherically-symmetric 2-body problem considered in  Sect. \ref{Sect:2body}, where there are no unstable orbits\footnote{This changes in general relativity, where even the spherically-symmetric Schwarzschild spacetime admits unstable orbits \cite{Wald, PoissonWill}. The Newtonian problem also becomes unstable if the number of spatial dimensions is increased from three to at least five.} for spherically-symmetric satellites. Stabilization is more relevant to the 3-body problem, where one might attempt to maintain position around an ordinarily-unstable Lagrange point, or to counteract the Moon-induced eccentricity increase \cite{Beletsky, stableSat}---and possible crash---of a satellite which is in low polar orbit around the Earth. 

We do not enter here into the complexities of 3-body stabilization or station-keeping, but instead introduce a model which illustrates the relevant principles. This model consists of a single gravitating body, perhaps an oblate planet or moon, which is symmetric only with respect to rotations around one axis. If the axis of symmetry is tangent to the unit vector $\hat{\bm{n}}$, we suppose more specifically that the potential at a spacecraft's center of mass $\bz$ is 
\begin{align}
	\Phi = - \frac{ GM }{ r} \left[ 1 + \frac{ J }{ 2r^2 } \left(  1 - 3 ( \hat{\bz} \cdot \hat{\bm{n}} )^2 \right) \right],
	\label{phiJ}
\end{align}
where $J$ is a constant proportional to the quadrupole moment of the central mass. Although this satisfies $\nabla^2 \Phi = 0$, it is an idealization. The potential of a realistic oblate body would typically involve these terms together with corrections which fall off with larger inverse powers of $r$ \cite{PoissonWill}. Such corrections are omitted here for brevity, although including them would not change anything essential in the analysis below.

Motion in the potential \eqref{phiJ} is in general quite complicated, even in the absence of extended-body effects. Nevertheless, considerable simplifications arise if we restrict to the equatorial plane orthogonal to $\hat{\bm{n}}$. Doing so while also assuming that $J>0$ and that the central mass is sufficiently compact, there exist unstable circular orbits for spherically-symmetric test masses. More precisely, if $(L/m)^4 > 6 J (GM)^2$, the unstable circular orbits occur at
\begin{equation}
	\rc = \frac{ (L/m)^2 }{ 2 G M } \left[ 1 - \sqrt{ 1 - \frac{ 6 J (GM)^2 }{ (L/m)^4 } } \right].
\end{equation}
These orbits satisfy\footnote{For the Earth, $J = 10^{-3} R_\oplus^2$ \cite{stableSat}. Formally applying the criterion for unstable circular orbits thus results in $\rc < 0.04 R_\oplus$, which is impossible. This calculation nevertheless illustrates that oblateness can induce unstable circular orbits only for bodies which are very compact or very non-spherical. The Earth is neither.} $\rc < \sqrt{3J/2}$. If a spherical spacecraft is initially on a circular orbit with such a radius, almost any slight perturbation would result in it spiraling either towards the central body or away from it. The remainder of this section establishes that such instabilities can be eliminated using extended-body effects.

To simplify the calculations, again suppose that a spacecraft is torque-free. Generalizing the discussion of Sect. \ref{Sect:torquefree}, it is useful for this purpose to assume that both the radial vector $\bm{z}$ and the symmetry vector $\hat{\bm{n}}$ are eigenvectors of a spacecraft's quadrupole moment. The radial eigenvector condition \eqref{eigQ} is thus supplemented by 
\begin{equation}
	Q^{ij} \hat{n}_j = q' \hat{n}^i
\end{equation}
for some eigenvalue $q'$. Orbits may be controlled by varying $q$ and $q'$. If an orbiting spacecraft is axisymmetric with an axis of symmetry tangent to that of the central body, $q$ must be a degenerate eigenvalue and the trace-free nature of $Q^{ij}$ implies that $q' = - 2 q$. More generally, there is not necessarily any relation between $q$ and $q'$. The two eigenvector assumptions are nevertheless sufficient to ensure that if a spacecraft is initially moving in the equatorial plane of the central body, it remains in that plane. These assumptions also ensure that $\bm{\tau} = \bm{0}$ for equatorial orbits. The discussion at the end of Sect. \ref{Sect:ConsLaws} implies that $L = |\bL| = |\bL \cdot \hat{\bm{n}}| = \mathrm{constant}$ in this context.

It follows from \eqref{Ediff} that for a spacecraft moving in the equatorial plane and satisfying the two aforementioned eigenvector conditions, the energy can be written in the form $E = \frac{1}{2} m \dot{r}^2 + U_\mathrm{eff}(r,q,q')$, but now the effective potential energy \eqref{Ueff} is replaced with
\begin{align}
	U_\mathrm{eff}(r, q, q') = \frac{L^2}{2 m r^2} - \frac{ GMm}{r } \bigg[ 1 + \frac{ J }{ 2 r^2 } - \frac{ 3 J q' }{ 2 m r^4 }  
	\nonumber
	\\
	~ +  \frac{3q}{2mr^2}  \left( 1+ \frac{ 5 J}{2 r^2 } \right) \bigg].
\end{align}
Eq. \eqref{QdotZero} implies that $E$ is constant whenever $q$ and $q'$ are themselves constant. If $q = q' =0$, the effective potential admits a local maximum at $r=\rc$.

\begin{figure}
    \centering
    \includegraphics[width=1\linewidth]{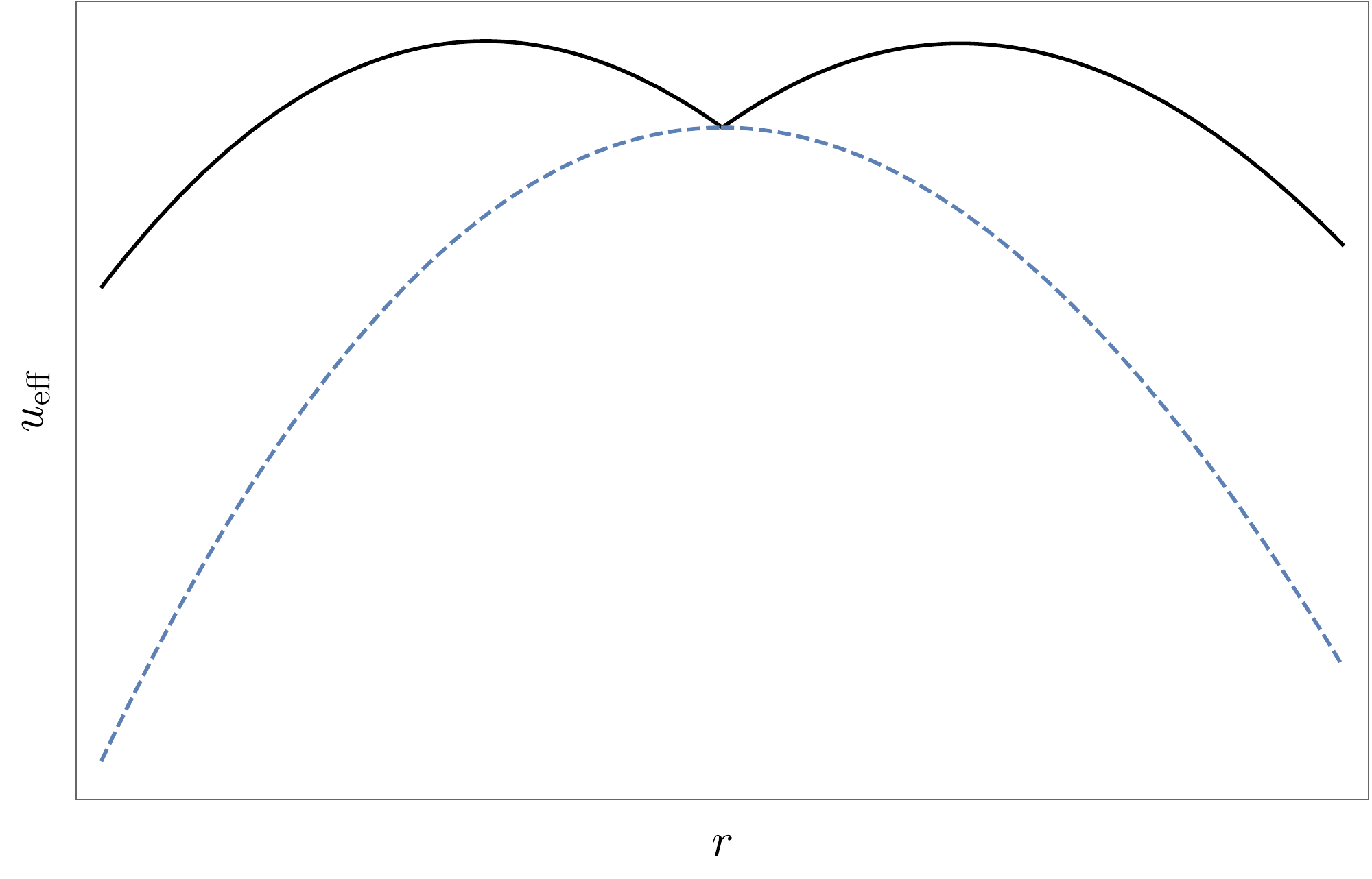}
    
    \caption{Stabilization of an unstable orbit by quadrupole modulation. The solid line is $u_\mathrm{eff}$ with $q$ and $q'$ given by \eqref{qSwitch} and $q_- = - q_+$. The dashed line is a vertical translation of the point-particle effective potential $U_{\mathrm{eff}}(r,0,0)$. The cusp is at $r = \rc$ and a body placed near that radius remains nearby as long as its energy is not too large.}
    \label{Fig:Unstable}
\end{figure}

Extended-body effects may be used to stabilize a circular orbit by varying the quadrupole moment such that the magnitude of the gravitational force increases when $r > \rc$ and decreases when $r < \rc$. This can be achieved by holding $q$ and $q'$ constant except when the spacecraft crosses the radius $\rc$. Suppose for definiteness that there are constants $q_\pm$ such that
\begin{equation}
	q = - \tfrac{1}{2} q' = q_+ \Theta( r - \rc ) + q_- \Theta( \rc - r).
	\label{qSwitch}
\end{equation}
Unlike in Sect. \ref{Sect:2body}, it is necessary for the strategy described here to demand that both $q_+$ and $q_-$ be nonzero and that they have opposite signs. Regardless, $E$ has one value, say $E_-$, when $r < \rc$. It has another value, $E_+$, when $r > \rc$. These values remain the same no matter how many times the spacecraft crosses the $r=\rc$ sphere. Furthermore, \eqref{dEGen} implies that $E_+$ and $E_-$ are related via
\begin{equation}
    E_+ - E_- =  -\frac{ 3 GM  }{ 2 \rc^3 } ( 1 +9J/2r^2 ) ( q_+ - q_- ) .
\end{equation}
This difference is always negative. The main result\footnote{An equivalent relation can be written down in which $E_+$ is emphasized instead of $E_-$.} is now that \textit{for all} $r$,
\begin{equation}	
	E_- = \tfrac{1}{2} m \dot{r}^2 + u_\mathrm{eff},
	\label{Eminus}
\end{equation}
where the effective potential which appears here is given by
\begin{align}
     u_\mathrm{eff} = U_\mathrm{eff}(r,q,-2q) - (E_+ - E_- ) \Theta(r- \rc).
     \label{ueffUnst}
\end{align}
The radial motion of a spacecraft may be understood directly from $u_\mathrm{eff}$. This function is continuous but not differentiable at $r = \rc$. It is plotted schematically in Fig. \ref{Fig:Unstable} for the case $q_- = - q_+$, where it may be seen that extended-body effects carve out a small well on top of the approximately-parabolic point-particle effective potential. An object in this well is manifestly stable. This may be understood somewhat more intuitively by noting that if $q = -\frac{1}{2} q'$ were constant and nonzero, the peak of the effective potential would be shifted away from $r = \rc$. A quadrupole moment which instead switches according to \eqref{qSwitch} shifts the peak in one direction when $r < \rc$ and in the opposite direction otherwise. As a consequence, the single peak which arises in the absence of extended-body effects is split into two closely-separated peaks. These are separated by a valley which stabilizes the system.

\section{Discussion}

Spacecraft can control their orbits by controlling their shapes. Roughly speaking, any maneuver which is not forbidden is allowed, and rocket-free maneuvers can be forbidden only by the symmetries of the gravitational field\footnote{This refers to what is possible in principle. There may be practical considerations which provide additional constraints, such as the limitation on spin discussed in Sect. \ref{Sect:torquefree}.}. We have focused on how such effects can be understood using the general theory of extended-body motion. It has been emphasized that all relevant details of a spacecraft's interior can be described by its quadrupole moment, which obviates the need for detailed interior models.

This perspective has been applied to two particular problems: torque-free orbital maneuvering around a spherical mass and torque-free station-keeping around a highly-oblate mass. Our focus on torque-free configurations is intended to i) avoid the development of excessively-large spins, and ii) to fix attention on an easily-analyzed class of spacecraft. We have shown that the torque vanishes when the quadrupole moment possesses particular eigenvectors, and in these cases, the associated eigenvalues can be varied to control an orbit. One possible generalization of the work considered here would be to understand if more efficient maneuvers could be executed by spacecraft in which it is only the averaged torque which vanishes.

Even without such generalizations, our analysis allows new questions to be answered. How, for example, can a spacecraft efficiently alter the eccentricity of its orbit without also altering the apsidal orientation of that orbit? In a spherically-symmetric gravitational field and in the torque-free case for which the eccentricity change is optimal, it follows from \eqref{deAv} and \eqref{psiEcc} that this occurs when $q_+= -q_-$, where $q_\pm$ are the maximum and minimum values of the eigenvalue $q$ which is associated with the radial eigenvector of a spacecraft's quadrupole moment. Many spacecraft with this property may be designed, although they differ from those which have been considered in the existing literature. For the commonly-analyzed \cite{Beletsky, Landis, Longo} dumbbell-type spacecraft whose orientations are held fixed either perpendicular to the orbital plane or parallel to the local vertical, $q$ has a fixed sign and precession is inevitable. That sign can change if the spacecraft has a mechanism which allows it to switch between the orthogonal and vertical orientations, and in that case, precession-free eccentricity changes are possible.

We have focused here only on motion in Newtonian gravity. However, one of our original motivations was to better understand what is possible in Newtonian dynamics in order to contrast with certain extended-body effects which have been claimed to arise in general relativity. Describing an object's configuration in terms of its multipole moments is more critical in relativistic contexts, as it is difficult to construct explicit, physically-plausible models which are also tractable. Indeed, it can be difficult even to verify that certain models fall freely and do not suffer from negative energy densities or other pathologies. Difficulties of this kind have led \cite{schwGlide, VitekThesis} to certain claims \cite{Swimming} for qualitatively non-Newtonian extended-body effects in general relativity, claims which were in part refuted by methods related to the ones considered here \cite{RescueSwimming}.

Nevertheless, certain aspects of the relativistic problem are indeed different. For example, the momentum of a relativistic extended body is not necessarily proportional to its velocity, and the misalignment between these two vectors---associated with a body's ``hidden momentum'' \cite{Bobbing, Hidden}---can be controlled using internal deformations \cite{CosmSwimming}. Relativistic extended-body effects may thus be used to directly modulate not only a body's acceleration, but also its velocity. This is true for both gravitational and electromagnetic interactions.

\end{document}